\documentclass[pra,twocolumn,english,showpacs,floatfix]{revtex4-1}
\usepackage{graphicx}
\usepackage{amssymb}
\usepackage{amsmath}
\usepackage{color}
\usepackage{amsmath}
\usepackage{xfrac}
\usepackage{nicefrac}

\usepackage{adjustbox} 
\usepackage{tabularx}
\usepackage{booktabs}
\usepackage{multirow}
\usepackage{makecell}

\newcommand\aver[1]{\langle#1\rangle}
\def \ket#1{{|#1\rangle}}

\begin{document}
\title{Dissipative time evolution of a chiral state after a quantum quench }
\author{Stefan Wolff}
\author{Ameneh Sheikhan}
\author{Corinna Kollath}

\affiliation{HISKP, University of Bonn, Nussallee 14-16, 53115 Bonn, Germany}

\begin{abstract}
We investigate the dynamics of fermionic atoms in a high-finesse optical resonator after a sudden switch on of the coupling between the atoms and the cavity. The atoms are additionally confined by optical lattices to a ladder geometry. The tunneling mechanism on a rung of a ladder is induced by a cavity assisted Raman process. At long times after the quantum quench the arising steady state can carry a chiral current.  In this work we employ exact diagonalization techniques on small system sizes to study the dissipative attractor dynamics after the quench towards the steady state and deviations of the properties of the steady state from predictions obtained by adiabatically eliminating the cavity mode.
\end{abstract}

\date{\today}
\maketitle

\section{Introduction}
In recent years the effort devoted to the investigation of topologically non-trivial quantum states revived due to the discovery of topologically insulating materials \cite{HasanKane2010}.  The motivation stems from the belief that such states could be of great use due to the protection of the state by the topology. For example topological quantum computation has been proposed to circumvent the problem of decoherence of useful quantum states. Recently, topologically non-trivial states have been created in quantum gases for example by the application of strong artificial magnetic fields \cite{DalibardOehberg2011}. These act on neutral atoms similar to magnetic fields act on charged particles. Experimentally, the Hofstadter model in two dimensions \cite{JakschZoller2003, AidelsburgerBloch2013,MiyakeKetterle2013,AidelsburgerGoldman2015} or on a ladder geometry \cite{AtalaBloch2014} and the Haldane model \cite{JotsuEsslinger2014} have been realized in optical lattices.

The dynamics in topologically non-trivial models has attracted increased attention since the possibility to switch rapidly between topologically distinct phases in materials became within experimental reach. Theoretically, the dynamic control of such phases,  e.g.~by a quantum quench (see for example \cite{RahmaniChamon2010,Perfetto2013,HalaszHamma2013,TsomokosFazio2009,FosterYuzbashyan2013,WangXianlong2015,DongPu2014b,FosterYuzbashyan2014,PatelDutta2013,WangKehrein2016,DongPu2015,
DAlessioRigol2015,WangKehrein2015,CaioBhaseen2015} and citations therein), have been started to be investigated. Away from equilibrium, one of the difficulties is to define useful topological quantum numbers. This becomes evident at the example of a two-band Chern insulator in which the Chern number does not change during the dynamics, but the Hall conductance is logarithmically divergent \cite{WangKehrein2015}. Similarly, after a quantum quench in the Haldane model, the Chern number does not change, but the signatures of the edge state as a chiral current can develop \cite{CaioBhaseen2015}.

A different approach to protect quantum states makes use of tailored environments \cite{MuellerZoller2012}. Here, the environment is engineered in such a way, that the steady state is the targeted quantum state. Due to the dissipative influence of the environment an exponentially fast decay towards a so-called {\it attractor} state can take place. Any external perturbation will be followed by a decay towards the attractor state. Many interesting states have been proposed along these lines. Examples reach from Bose-Einstein condensation and BCS pairing to topologically non-trivial quantum states \cite{MuellerZoller2012}.

One experimental setup which has been experimentally realized in the past and in which the dissipative dynamics is of great importance is an atomic quantum gas placed into an optical high-finesse resonator \cite{RitschEsslinger2013}. This setup allowed the recent realization of an open-system version of the Dicke phase transition \cite{Dicke1954,HeppLieb1973, WangHioe1973}.  To this end, a transverse laser beam was applied and the organization of the atoms into a checkerboard density pattern was observed above a critical pump strength \cite{BaumannEsslinger2010,KlinderHemmerich2015,DomokosRitsch2002,NagyDomokos2008,RitschEsslinger2013,PiazzaZwerger2013,DimerCarmichael2007,BadenBarrett2014}. In this setup, different super-radiant fixed points~\cite{BhaseenKeeling2012,LiuJia2011} have been investigated theoretically. The additional application of external optical lattice potentials has been achieved~\cite{KlinderHemmerich2015b,LandigEsslinger2016} and the influence of the presence of the cavity-mediated long-range interactions onto the superfluid to Mott insulator phase transition has been investigated \cite{LarsonLewenstein2008, MaschlerRitsch2005, MaschlerRitsch2008, NiedenzuRitsch2010,SilverSimons2010,VidalMorigi2010, LiHofstetter2013,RitschEsslinger2013,BakhtiariThorwart2015}. 

Theoretically, further examples of similar systems have been put forward as bosonic atoms organize into triangular or hexagonal lattices \cite{SafaeiGremaud2015} or fermionic atoms driven into super-radiant phases \cite{LarsonLewenstein2008b,MuellerSachdev2012,PiazzaStrack2014,KeelingSimons2014,ChenZhai2014}. Also phases with spin-orbit coupling have been proposed in standing-wave cavities \cite{DengYi2014,DongPu2014,PanGuo2015,PadhiGosh2014} or ring cavities \cite{MivehvarFeder2014,MivehvarFeder2015}. 

Whereas the steady states are the subject of intensive studies, much less is known about the dynamics in the coupled cavity atom systems. First investigations of dynamic correlations \cite{KulkarniTuereci2013}, the damping of quasi-particles \cite{KonyaDomokos2014}, self-ordered limit cycles \cite{PiazzaRitsch2015}, or prethermalization effects \cite{SchuetzMorigi2014} have been performed theoretically.

In this work we will describe how fermionic atoms which are suddenly coupled to an optical cavity mode reach via the feedback mechanism of the cavity a steady state which carries a chiral current. The work extends our previous work \cite{KollathBrennecke2016,SheikhanKollath2016} in which mainly the steady state diagram of such a system has been investigated. The focus of the present work is the discussion of the dynamics after the quench of the coupling towards the steady state. The dynamics resembles that of an attractor dynamics and might therefore enable a fast and stable preparation of such chiral current carrying states.

In Sec.~\ref{sec:model} we introduce the model describing the atoms in the optical cavity. In Sec.~\ref{sec:methods} we give a detailed description of the applied methods. The first method is an approximative analytic approach using the adiabatic elimination of the cavity field, which is suitable for steady state predictions. The second approach is based on numerically exact simulations using a full diagonalization of the Lindblad superoperator. This approach provides not only information on the steady states, but also the time-evolution of the open system. We present results on the properties of the steady states of the system in Sec.~\ref{sec:steady_state}, including a steady state diagram where a chiral liquid phase arises. In Sec.~\ref{sec:dynamics} we focus on the relaxation dynamics after the coupling quench, before we discuss our results in Sec.~\ref{sec:discussion}.
 
\section{Model}
\label{sec:model}
\begin{figure}[hbtp]
\centering
\includegraphics[width=.5\textwidth]{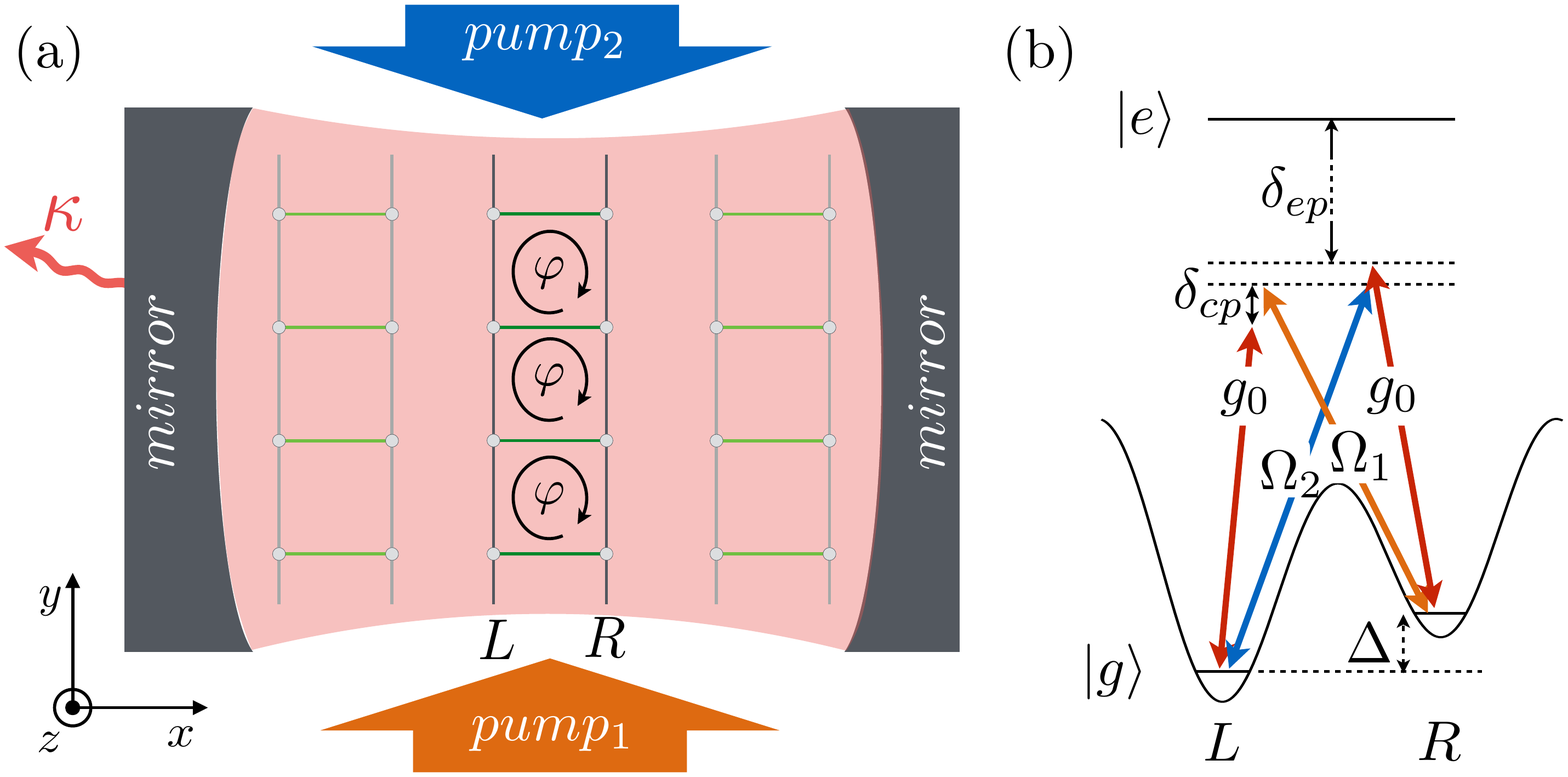}
\caption{(a) Sketch of the set-up (see text). The optical lattice beams forming the ladder geometry for the atoms are not shown.  (b) Schematic level scheme and balanced Raman processes. $\Omega_{1,2}$ and $g_0$ represent the Rabi frequencies of the two  pump beams and the cavity, respectively. The energy off-set between $L$ (left) and $R$ (right) sites along the $x$-direction is denoted by $\Delta$. The pump beams and the cavity field are assumed to be far detuned from the atomic transition $\ket{g}\rightarrow \ket{e}$.}
\label{fig:set-up}
\end{figure}

We consider ultracold spinless, non-interacting fermionic atoms placed in an optical Fabry-Perot cavity. The atoms are trapped in an optical lattice potential, formed by orthogonal standing wave laser beams, engineered such that decoupled ladder structures arise (Fig.~\ref{fig:set-up} a). Along the $z$-direction a strong optical lattice potential is applied which confines the atoms into pancake like structures.
Within the $x$-$y$-plane a lattice potential with wavelength $\lambda_y$ is applied along the $y$-direction and a superlattice combination with wavelengths $\lambda_x$ and $\lambda_x/2$ along the $x$-direction. This superlattice structure is arranged such that almost decoupled double wells are formed along $x$ with a potential difference of $\Delta$ between the two sites labeled $L$ and $R$ (see Fig.~\ref{fig:set-up} b). Whereas tunneling along the $y$-direction occurs with amplitude $J_\|$, tunneling along the rungs is strongly suppressed. It is restored by Raman transitions involving two running-wave pump laser beams and a standing wave cavity mode. The counter propagating pump beams are applied transversely to the cavity direction and have frequencies $\omega_{p,i=1,2}$. The frequency of the first pump laser beam is tuned as $\hbar(\omega_{p,1}-\tilde{\omega}_c) \approx \Delta$ and the frequency difference of the two pump beams is given by $\omega_{p,2}-\omega_{p,1}=2\Delta/\hbar$. Here, the dispersively shifted resonance frequency of the cavity mode is denoted by $\tilde {\omega}_c$. The inelastic scattering of the atoms gives rise to a balanced pair of two-photon Raman transitions via one pump beam and the cavity mode (see Fig.~\ref{fig:set-up} b). These processes create or destroy photons in the cavity mode together with a hopping and therefore, establish a feedback mechanism between the cavity field and the motion of the atoms. The emerging lattice geometry yields a set of decoupled ladders with ordinary tunneling along the leg direction and photo-induced tunneling on the rungs (see Fig.~\ref{fig:set-up} a).

The running-wave character of the transverse pump beams imprints a phase factor $\text{e}^{i \bf{k\cdot r}}$  onto the tunneling process along the rungs. Here, the wave-vector ${\bf k}$ denotes the difference between the cavity $k_c {\bf e}_x$ and the pump wave vector $k_{p,i=1,2} {\bf e}_y$. As the cavity-assisted tunneling is directed along the $x$-direction, circulating around a plaquet of the ladder induces a finite phase change which arises from the transverse pump beam, while the phases mediated by the horizontal cavity field compensate each other. More precisely, a space-dependent phase of $j\varphi$ with $\varphi\equiv k_p a_y$ is imprinted by the pump beams onto the tunneling along the rungs, where $a_y$ is the lattice spacing and $j$ the site label along the legs of the ladders. We made use of the fact that the wavelengths of the pump beams are similar $\lambda_{p,1}\approx \lambda_{p,2} = \lambda_p$. Experimentally, the phase imprint $\varphi$ can be varied by tilting the pump beams out of the $x$-$y$-plane. We focus in the following on the situation $\varphi=\pi/2$.

For a large detuning of the Raman beams and the cavity modes from the transition frequency of the atoms $\omega_e$, the excited state is only sparsely populated and can be adiabatically eliminated. For convenience we reference to a frame rotating with frequency $\omega_p = (\omega_{p,1}+\omega_{p,2})/2$ and apply the rotating wave approximation. 

We expand the atomic field operators in the Wannier basis. Choosing sufficiently strong optical lattices guarantees well-localized atomic wave-functions and allows the restriction  of the fermionic part of the Hamiltonian to almost local terms. The resulting Hamiltonian describing the coupled atom-cavity system reads
\begin{align}
\label{eq:Hamiltonian}
&H=H_c+H_\parallel+H_{ac} \\
&H_c= \hbar\delta_{cp} a^\dagger a\nonumber\\
&H_\parallel=-J_\parallel \sum_{j,m=0,1} (c_{m,j}^\dagger c_{m,j+1} + c_{m,j+1}^\dagger c_{m,j})\nonumber\\
&H_{ac}= -\hbar\tilde{\Omega} ( a + a^\dagger) ( K_\perp + K_\perp^\dagger)\nonumber\\
&K_\perp=  \sum_{j=0}^{L-1} e^{i\varphi j}c_{0,j}^\dagger c_{1,j}\nonumber.
\end{align}

Here $c_{m,j}$ and $c_{m,j}^\dagger$ are the fermionic annihilation and creation operators of the atoms and $m$ labels the ladder legs and $j$ the ladder rungs, respectively. The number of rungs of the ladder is $L$. The mode of the cavity field is represented in the rotating frame by the bosonic operators $a$ and $a^\dagger$, while the pump beams are treated classically. The cavity contribution $H_c$ to the Hamiltonian in the rotating frame is proportional to $\delta_{cp}=\tilde{\omega}_c-\omega_p$. The tunneling along the ladder legs is expressed by $H_\|$ with the amplitude $J_\|$. The restored tunneling perpendicular to the legs denoted by $K_\perp$, is coupled to the cavity field operators. The balanced Raman transition scheme involving two pump beams has been introduced in order to couple each direction of the tunneling to  the creation or the annihilation of a cavity photon \cite{DimerCarmichael2007}. This prevents a privileged direction of the tunneling, which would occur in the case of a single pump beam and cavity loss. The effective amplitude of the cavity-assisted tunneling along the rungs induced by the first pump beam is given by $\hbar\tilde{\Omega}=\frac{\hbar\Omega_{p,1} g_0}{\omega_e-\omega_{p,1}} \phi_\| \phi_\perp$, where $g_0$ is the cavity Rabi frequency and $\phi_\|$ and $\phi_\perp$ are related to the overlap of neighboring site wave functions and are adjustable by the lattice geometry \cite{KollathBrennecke2016}. To provide similar effective amplitudes of the two Raman processes initiated by the two pump beams, the Rabi frequency of the second pump is chosen to be $\Omega_{p,2}= \Omega_{p,1}\frac{\omega_e-\omega_{p,2}}{\omega_e-\omega_{p,1}}$.

In the case of a finite occupation of the dynamic cavity field,  the phase $\varphi$ imprinted during the tunneling around a plaquet acts as an artificial magnetic field felt by the fermionic atoms. This field can induce a chiral particle current $J_c$ enclosing the ladder, defined by 
\begin{equation}
J_c = \frac{1}{L-1} \sum_j \left(j_{0,j} - j_{1,j} \right)
\end{equation}
with $j_{m,j} = -i J_\| (c_{m,j}^\dagger c_{m,j+1} -\text{h.c.}) $ the current between sites $j$ and $j+1$ of the leg $m$. 
	
The imperfection of the cavity mirrors in an experimental realization gives rise to photon losses which introduce a dissipative nature to the system. The fact that a later retrospective action of the lost photons on the combined photon-atom coupled system can be neglected, allows the treatment of the system by a Markovian master equation of Lindblad form 
\begin{align}\label{eq:Lindblad}
& \dot \rho(t) = \mathcal{L}\rho = -\frac{i}{\hbar} \left[ H, \rho(t) \right] + \mathcal{D}(\rho(t)) \\
&\text{with }\mathcal{D}(\rho) =  \kappa \left( 2a \rho a^\dagger - \rho a^\dagger a - a^\dagger a \rho  \right). \notag
\end{align}
Here $\rho$ denotes the density matrix of the atoms and the cavity mode, the cavity field annihilation operator $a$ is the Lindblad jump operator, and $H$ is the  Hamiltonian introduced in Eq. (\ref{eq:Hamiltonian}). The Lindblad superoperator $\mathcal{L}$ generates a quantum dynamical semigroup $\{V(t)\vert t\ge 0\}$, represented by the linear map $V(t)= \exp(\mathcal{L} t)$. The time evolution of the combined atom cavity system is then described by $\rho(t)=V(t)\rho_0$ according to the initial state $\rho_0$. The dynamics conserves the trace ($\text{tr}(\rho) =1$) and respects the semi-positivity of $\rho$. The initial state flows towards the stationary state(s) which is (are) the fixed point(s) of the dynamical semigroup.

In our work we investigate the time evolution of a fermionic state which is suddenly coupled to the cavity mode. The initial state of the fermions is chosen to be a pure state and the cavity mode is initially set to be empty. At time $t=0$ the pump laser is switched on which induces the coupling between the atoms and the cavity. 

\section{Methods}
\label{sec:methods}
In this section we introduce two different methods to cope with the master equation (Eq.~\ref{eq:Lindblad}).
The first approach (Sec.~\ref{sec:adiabatic}) is the adiabatic elimination of the cavity field from the equations of motion. The resulting effective model for the atoms needs to be considered with a self consistent equation for the cavity field and can be solved exactly. The second approach (Sec.~\ref{sec:ED}) solves the full Lindblad master equation for very small system sizes numerically and is well suited to describe the time evolution of the system after a quench.
  
\subsection{Adiabatic elimination of the cavity field}
\label{sec:adiabatic}
In this section we describe an approximation which is often performed, the adiabatic elimination of the cavity field. Since the cavity field evolves much faster than the atoms move, the cavity field can be assumed to take instantaneously its steady state value. This means that one can assume that the stationary state condition $\partial_t \langle a\rangle=0$ is fulfilled for the cavity field. Thus, the  equation of motion for the expectation value of the cavity field derived from Eq.~(\ref{eq:Lindblad}) becomes
\begin{equation}\label{eq:dyn_photon}
i \partial_t \langle a\rangle=0=-\tilde{\Omega}\langle K_\perp + K_\perp^\dagger \rangle +(\delta_{cp}- i \kappa ) \langle a\rangle.
\end{equation}
This leads to a relation between the cavity field and the fermionic rung tunneling given by $$\alpha=\langle a \rangle = \frac{\tilde{\Omega}}{\delta_{cp}- i \kappa }\langle K_\perp + K_\perp^\dagger \rangle. $$ 
One can substitute this expectation value of the cavity field operator into the equations of motion for the atomic operators assuming a mean-field decoupling of the photonic and atomic degrees of freedom. 
The equations of the fermionic operators become
\begin{align}\label{eq:dyn_fermion}
&i \hbar \partial_t c_{0,j} = -J_\parallel (c_{0,j+1} + c_{0,j-1})-\hbar \tilde{\Omega}\aver{a + a^\dagger} e^{i\varphi j} c_{1,j}\nonumber\\
&=-J_\parallel (c_{0,j+1} + c_{0,j-1})-J_\perp e^{i\varphi j} c_{1,j}\nonumber \\
&i \hbar \partial_t c_{1,j}= -J_\parallel (c_{1,j+1} + c_{1,j-1})-\hbar \tilde{\Omega}\aver{a + a^\dagger} e^{i\varphi j} c_{0,j}\nonumber \\
&= -J_\parallel (c_{1,j+1} + c_{1,j-1})-J_\perp e^{i\varphi j} c_{0,j}
\end{align}

with the self consistency condition $J_\perp=A\langle K_\perp \rangle$ and the proportionality constant $A= \frac{4\hbar\tilde{\Omega}^2\delta_{cp}}{\delta_{cp}^2 +\kappa^2}$. Here we make use of the fact that $\aver{K_\perp}$ is real. 

The effective Hamiltonian which corresponds to these equations of motion is
\begin{align}
&H_F=H_\parallel+H_\perp \label{eq:eff_ham}
\\
&H_\parallel=-J _\parallel \sum_{j,m=0,1} (c_{m,j}^\dagger c_{m,j+1} + c_{m,j+1}^\dagger c_{m,j})\nonumber\\
&H_\perp= -J_\perp (K_\perp + K_\perp ^\dagger) \nonumber
\end{align}
This Hamiltonian describes charged fermionic particles (non-interacting) subjected to a magnetic field. In the considered situation the atoms are neutral and the Raman transitions induced by pump and cavity fields leads to an artificial magnetic field \cite{JakschZoller2003, AidelsburgerBloch2013,MiyakeKetterle2013,AidelsburgerGoldman2015,DalibardOehberg2011}.
The particularity of our setup is the feedback mechanism between the cavity field and the atomic motion which leads to a dynamic occupation of the cavity field. Thus, the atoms are subject to a dynamically organized artificial magnetic field.

Since the effective Hamiltonian (Eq. \ref{eq:eff_ham}) is quadratic in the fermionic operators, it can be diagonalized by a Bogoliubov transformation \cite{CarrNersesyan2006,RouxPoilblanc2007,JaefariFradkin2012,HuegelParedes2014,TokunoGeorges2014}. The self-consistent solution can be obtained and has been discussed in detail in Refs.~\cite{KollathBrennecke2016,SheikhanKollath2016}.

\subsection{Exact diagonalization approach}
\label{sec:ED}
To provide an independent analysis of the steady states and to access the dynamical properties of the coupled cavity-atom system, we complement the above approach (Sec.~\ref{sec:adiabatic}) by numerically solving the Lindblad master equation (Eq.~\ref{eq:Lindblad}) for small system sizes. 
As $\mathcal{L}$ is a linear map, we can rewrite its action on the density matrices as a matrix $M_\mathcal{L}\in \mathbb{C}^{D^2\times D^2}$ 
\begin{equation}
\frac{\partial}{\partial t}\vert\rho(t)\rangle=M_\mathcal{L}\vert\rho(t)\rangle.
\end{equation}
Here, the density matrix $\rho\in \mathbb{C}^{D\times D}$ is transformed to a vector $\vert \rho \rangle \in \mathbb{C}^{D^2}$ where $D$ is the dimension of the system's Hilbert space. In contrast to a Hamiltonian matrix describing the unitary evolution, the matrix $M_\mathcal{L}$ does not need to be Hermitian.  

We numerically determine the full eigensystem of the Lindblad matrix $M_\mathcal{L}$. The time-evolved density matrix is then a decomposition of the right (left) eigenvectors $\vert \nu \rangle$ ($\langle \nu\vert$),
\begin{equation}
\vert\rho(t)\rangle = \sum_\nu \text{e}^{\lambda_\nu t } \vert \nu \rangle\langle \nu \vert \rho_0\rangle .
\end{equation}
The real part of the eigenvalues $\lambda_\nu$ is always smaller or equal to zero due to the semi-positivity of the map. A finite negative real part of the eigenvalue induces an exponential decay in time. The only steady states of the evolution are those associated with vanishing eigenvalues. In the case of uniqueness of such a steady state the state acts as an attractor in time and the system approaches this state independently of the initial conditions.  In the case of the existence of several steady states,  i.e. the vanishing eigenvalues are degenerate, the associated eigenstates span a manifold which may contain non-physical density matrices, i.e.~density matrices $\rho$ which violate one of the conditions 
\begin{align}
&\,\,(i)\,\,\quad\rho = \rho^\dagger, &\notag\\ 
&\,(ii)\,\quad\text{tr}(\rho) =1, \label{eq:physicalconditions}\\ 
&(iii) \quad\rho \text{ is semi-positive definite.} \notag
\end{align}
We will in the following use the symmetries of the model in order to extract the physical steady states \cite{Prosen2012,BucaProsen2012,Torres2014,AlbertJiang2014}. In particular, we use the symmetries in order to block diagonalize $M_\mathcal{L}$, such that there is a unique eigenvector with vanishing eigenvalue in each symmetry sector. This uniqueness guarantees that the eigenvector corresponds to a physical steady state. We will present the most important symmetries in the following subsection. For more than one particle the number of steady states increases as summarized in Table \ref{tab:numbersteadystates}.

\def\tabularxcolumn#1{m{#1}}
\begin{table}
\centering
\begin{tabularx}{0.45\textwidth}{lXccc}
\Xhline{3\arrayrulewidth}
$L\qquad$ &$\:\:$ lattice & $\qquad$filling $n\qquad$ & $\# \text{steady states}$   \\ \Xhline{2\arrayrulewidth}
1 & \includegraphics[scale=.05]{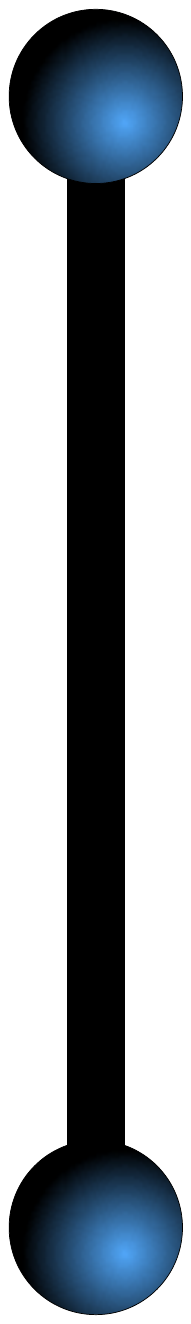} & 1/2  & 2\\ \hline
2 & \includegraphics[scale=.05]{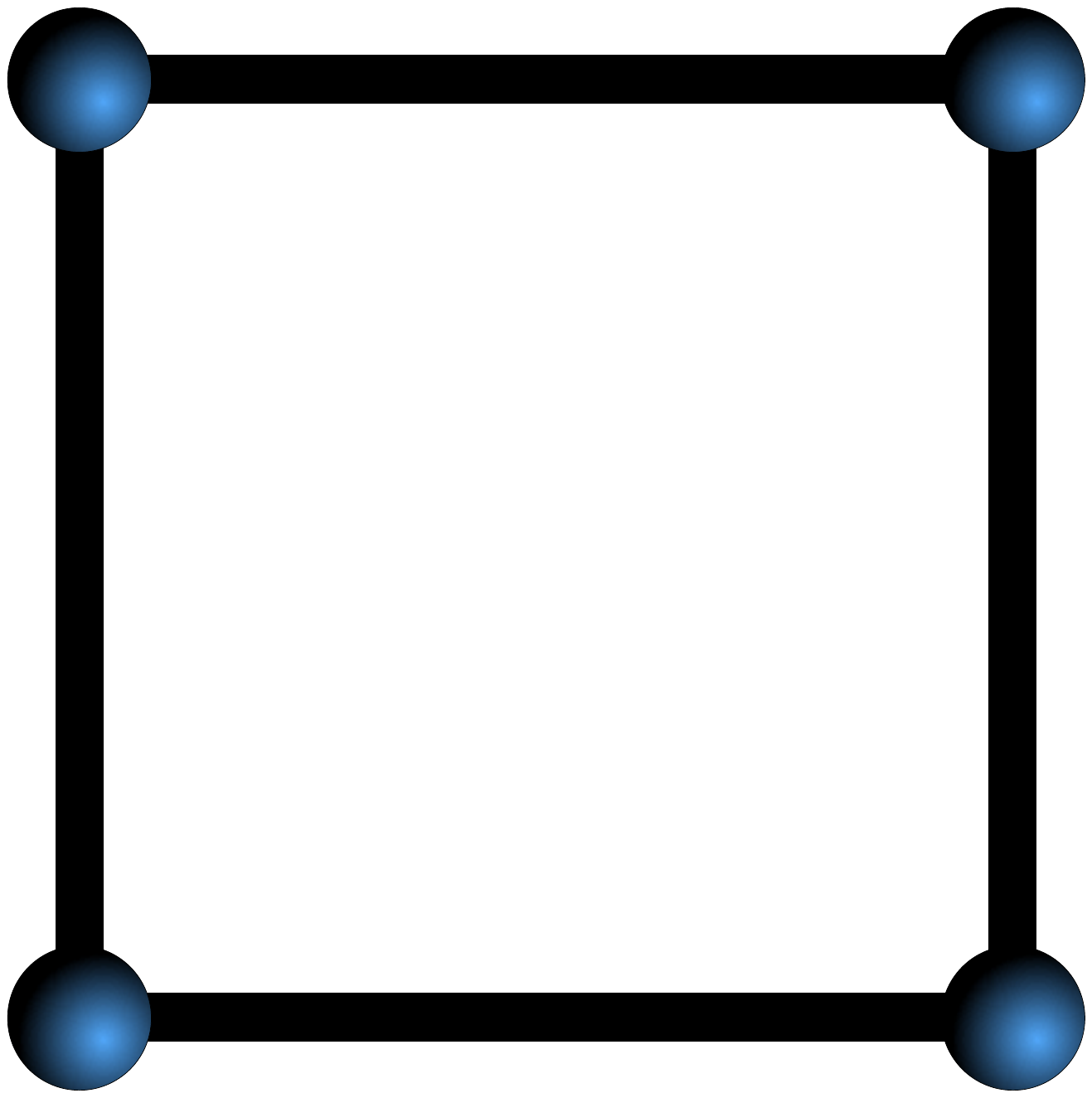} & 
\begin{tabular}{cc} 1/2 \\ 1/4 \end{tabular} & \begin{tabular}{cc} 6 \\ 2  \end{tabular} \\ \hline
3 & \includegraphics[scale=.05]{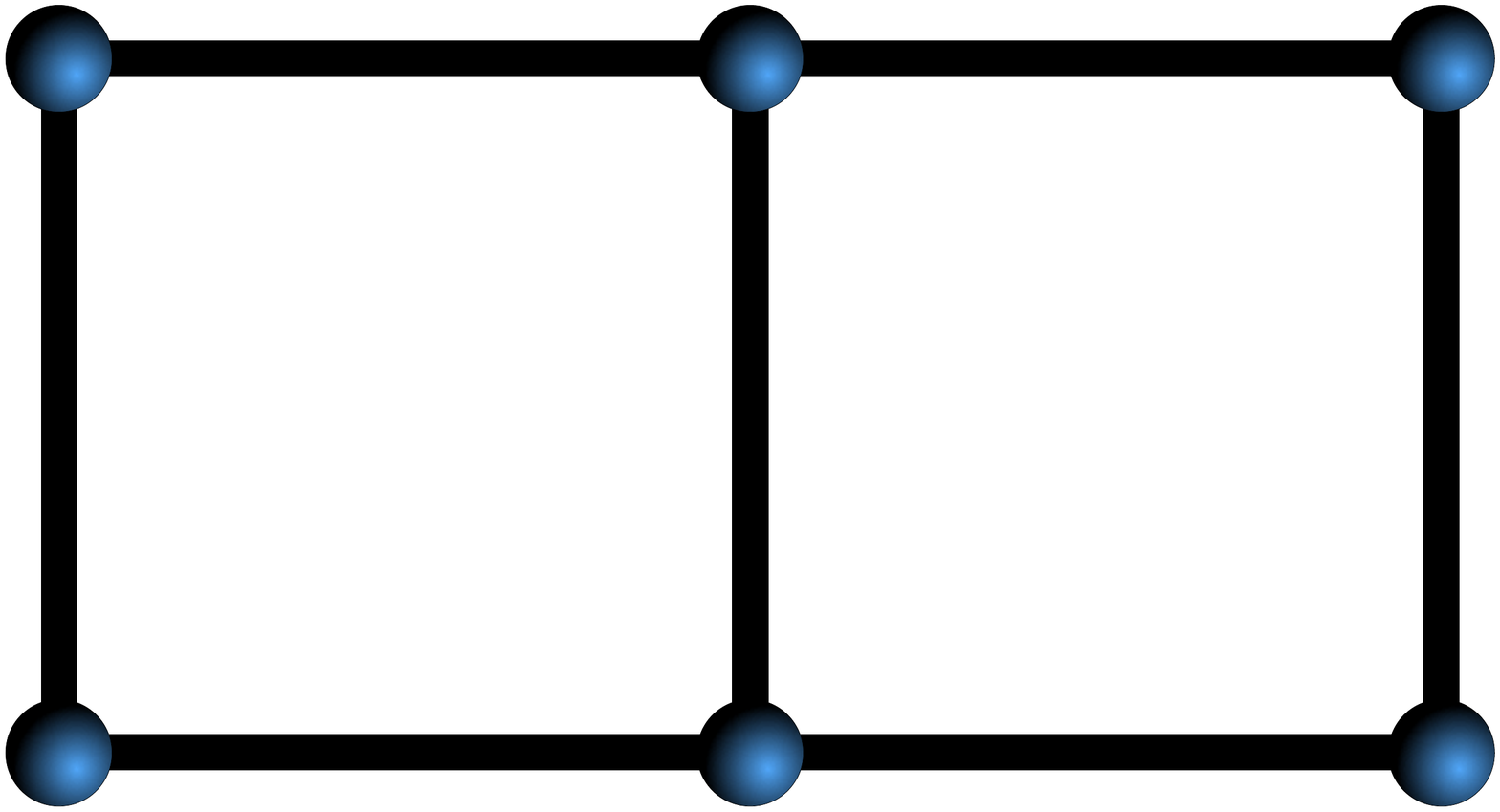} &\begin{tabular}{cc} 1/2 \\1/6\end{tabular}  & \begin{tabular}{cc} 6\\ 2\end{tabular}   \\\Xhline{3\arrayrulewidth}
\end{tabularx}
\caption{Number of steady states for different number of rungs of the ladder $L$, and different fillings $n$.}
\label{tab:numbersteadystates}
\end{table}

\paragraph{Symmetries:}
We discuss the symmetries which we use in the following for all considered geometries and fillings. 
The first symmetry we use is defined by the transformation $S_1$
\begin{align}\label{eq:symm}
&c_{0,j}\rightarrow e^{+i\frac{L-1}{2}\varphi} c_{1,L-1-j} \nonumber\\
&c_{1,j}\rightarrow e^{-i\frac{L-1}{2}\varphi} c_{0,L-1-j}.
\end{align}
The transformation $S_1$ leaves both $H_\parallel$ and $K_\perp+K_\perp^\dagger$ invariant. Consequently, also the full Lindblad master equation is invariant under $S_1$, since it only acts on the fermionic sector.  
This $\mathbb{Z}_2$-symmetry (Eq.~\ref{eq:symm}) enables us to write the fermionic part of the single particle Hamiltonian ($H_\parallel$ and $K_\perp+K_\perp^\dagger$) into two blocks. In each of the blocks a steady state exists and can be classified by the eigenvalues ($\pm 1$) of the symmetry transformation (Eq.~\ref{eq:symm}).

Moreover, the Hamiltonian (Eq.~\ref{eq:Hamiltonian}) possesses another $\mathbb{Z}_2$-symmetry given by the transformation $S_2$
\begin{eqnarray}\label{eq:symm_a}
\begin{cases}
    \,\,\, a\,\,\, \to -a\\
    c_{0,j}\to -c_{0,j}\\
    c_{1,j}\to \,\,\,\,c_{1,j}
\end{cases}
\Rightarrow\quad
\begin{cases}
    \,\,\,a\,\,\, \to -a\\
    K_\perp \to -K_\perp\\
    H_\parallel\, \to \,\,\,\,H_\parallel\, .
\end{cases}
\end{eqnarray}
This transformation also preserves the full Lindblad dynamics. $S_1$ and $S_2$ anticommute which means that the transformation $S_2$ maps any eigenvector of $S_1$ with the eigenvalue $+1$ in the first symmetry block onto a new eigenvector of $S_1$ corresponding to the eigenvalue $-1$ in the second symmetry block. Additionally, $K_\perp+K_\perp^\dagger$ and $S_2$ anticommute such that the transformation $S_2$ maps any eigenvector of $K_\perp+K_\perp^\dagger$ with eigenvalue $\lambda$ in the first symmetry block onto a new eigenvector of $K_\perp+K_\perp^\dagger$ corresponding to the eigenvalue $-\lambda$ in the second block. This means that steady states which lie in different symmetry blocks have different signs of the expectation value of $K_\perp+K_\perp^\dagger$. Therefore the cavity field expectation value of the two steady states in the different symmetry blocks will be related by a sign change, i.e.~$\pm \alpha$. We verified this behavior by our numerical explorations of the according symmetry blocks. In our mean field treatment the symmetry is broken by choosing the positive value for $\langle K_\perp\rangle$

\paragraph{Photon number restriction}

\begin{figure}[tp]
\centering
\includegraphics[width=0.5\textwidth]{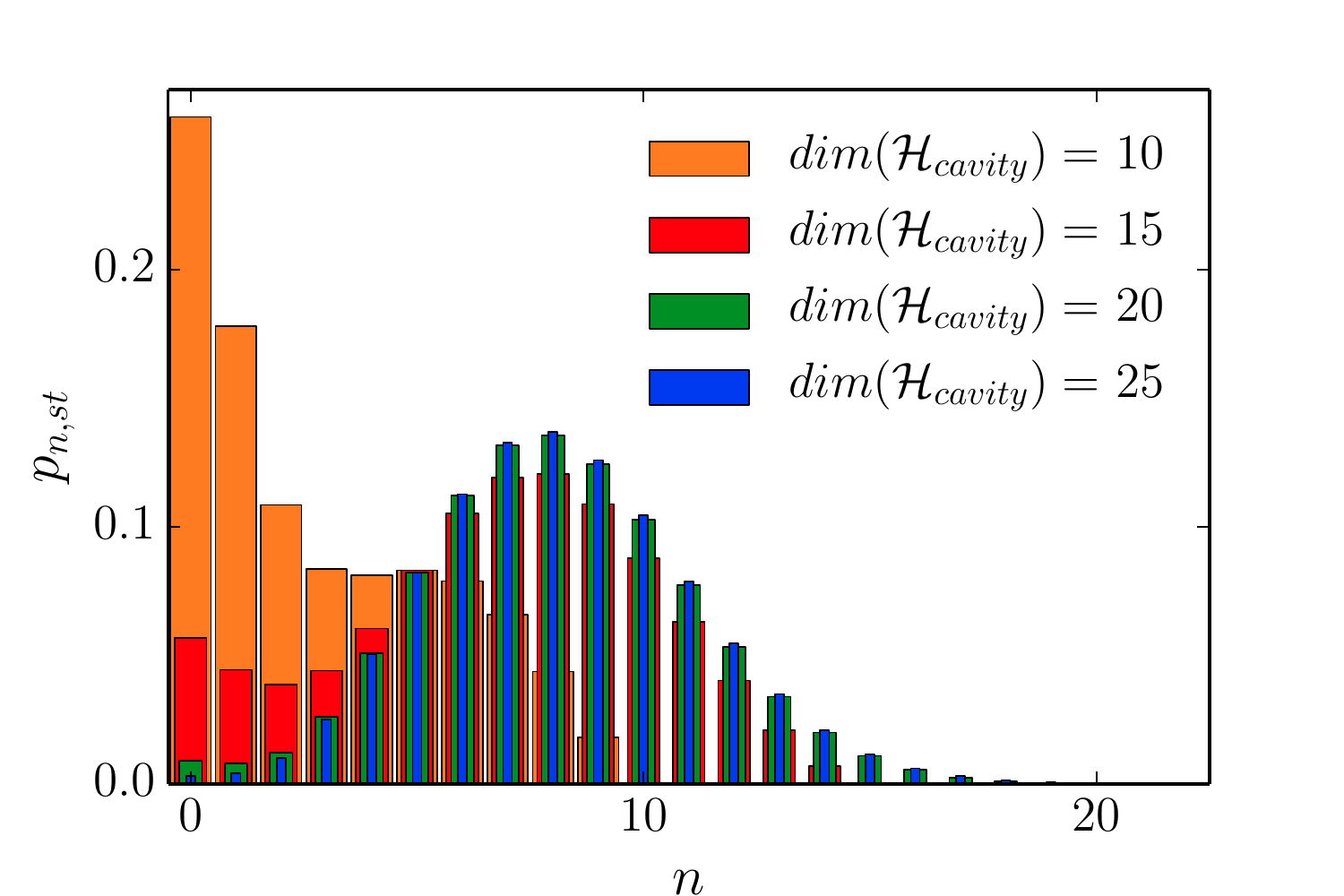}
\caption{Distribution of cavity photons in the steady state of a ladder with $L=3$ rungs, $N=3$ atoms, photon loss rate $\hbar\kappa/J_\|=0.25$, pump strength $\hbar \Omega/J_\| = 1.0$ and rotated cavity frequency $\hbar\delta_{cp}/J_\|=1.0$. The plot shows the result for the four different cutoffs of the bosonic Hilbert space up to $10$, $15$, $20$ and $25$ states.}
\label{fig:ConvergencePhotonDistributions}
\end{figure}

As the master equation for the studied model (Eq.~\ref{eq:Hamiltonian}) does not conserve the number of cavity photons, in principle an infinite-dimensional bosonic Fock space is required. However, in small systems with fermionic ladders up to three rungs and not to large pump amplitudes, the dominant part of the cavity field occupation in Fock space is found to be distributed in the region of low number states. Therefore, the photonic Fock space can be bounded by introducing a cutoff which defines a maximal number of photons populating the cavity mode. Fig.~\ref{fig:ConvergencePhotonDistributions} shows an example of the occupation of the bosonic Fock states in a steady state for different cutoffs. While smaller boson spaces which only take the $10$ or $15$ lowest number states into account suffer from errors due to the truncation of non-negligibly occupied states, the differences in the results for $20$ and $25$ number states are sufficiently small. We verified for all results shown below the convergence in the photon number cutoff. 

\section{Physical properties of the steady states}
\label{sec:steady_state}
\begin{figure}[hbtp]
\centering
\includegraphics[width=.5\textwidth]{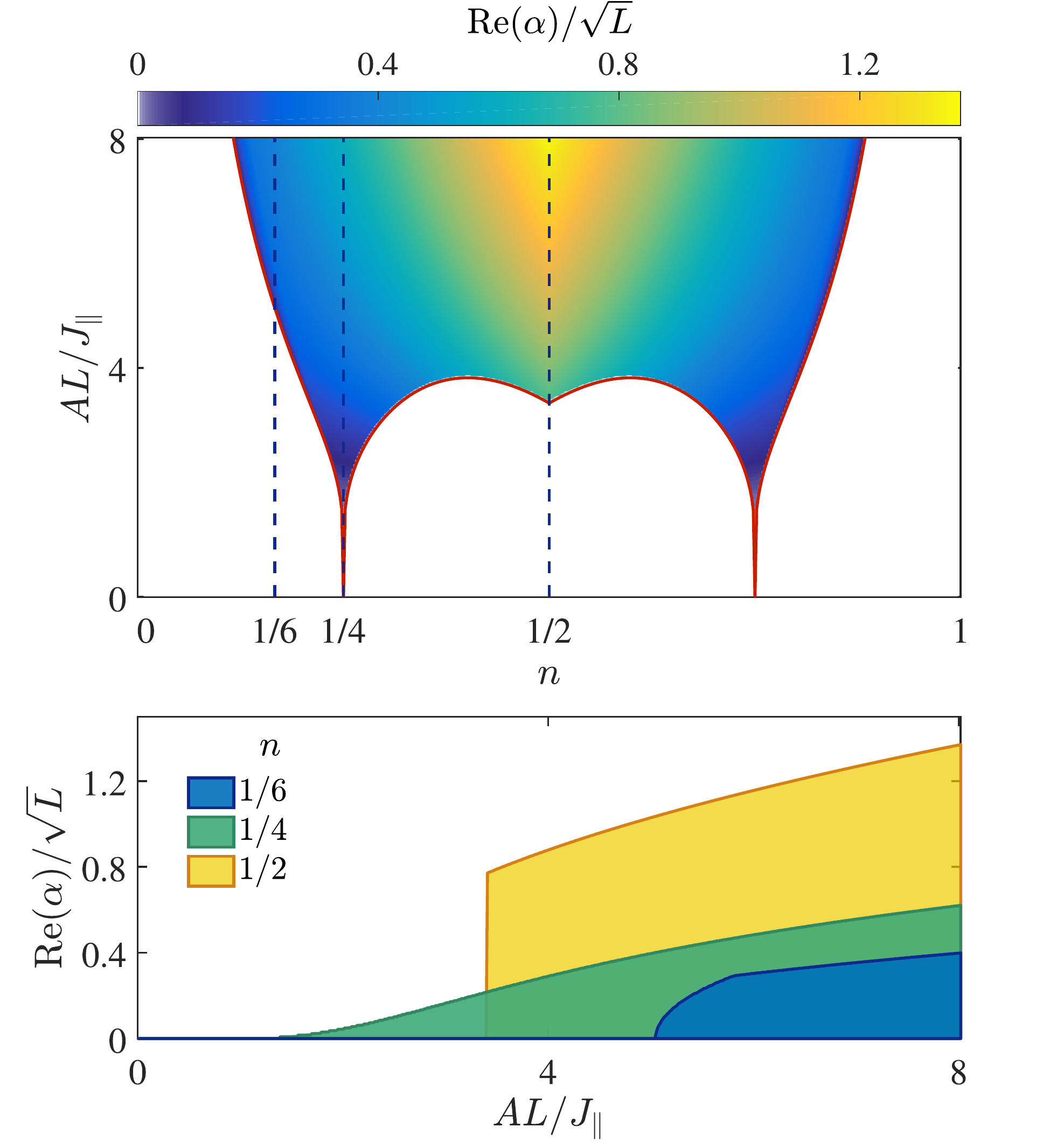}
\caption{Upper panel: Dependence of the real part of the rescaled cavity field on pump strength and atomic filling obtained by adiabatic elimination of the cavity field in the limit $L\rightarrow\infty$. The red curve denotes the critical value $A_{cr}$ and the dashed lines mark selected fillings shown in the lower panel. Lower panel: Cuts at different filling factors reveal the distinct behavior while passing the threshold between a vanishing field amplitude at low pump strength and finite occupation of the cavity mode at high pump strengths.}
\label{fig:alpha_n_A}
\end{figure}

In this section we discuss the properties of the steady states of the system extending previous results of references \cite{KollathBrennecke2016,SheikhanKollath2016}. In Fig.~\ref{fig:alpha_n_A} the real part of the cavity field is shown versus the atomic filling and the pump strength calculated using the adiabatic elimination of the cavity field.  This can be interpreted as an example of a steady state diagram. At most of the values of the atomic filling, a finite cavity field arises above a finite critical pump strength $A_{cr}$. The critical value of the pump strength $A_{cr}$ depends on the filling. Considering $\varphi=\frac{\pi}{2}$ we have \cite{SheikhanKollath2016}
\begin{align}
\left( \frac{AL}{J_{\|}}\right)_{cr} =
\begin{cases}
\frac{2\sqrt{2}\pi}{\log\left(\frac{\tan(\frac{\pi}{8}+\frac{n\pi}{2})}{\tan(\frac{\pi}{8}-\frac{n\pi}{2})}\right)}, 
& n < \frac{1}{4},n > \frac{3}{4}  \\
\frac{\left(\frac{J_\perp}{J_\parallel}\right)_{cr}}{K_\perp\left(\frac{J_\perp}{J_\parallel}\right)_{cr}/L}, 
&\frac{1}{4} < n< \frac{3}{4} .
\end{cases}
\end{align}
Further the equation for $(J_\perp)_{cr}$ is given by
\begin{equation}
\left(\frac{J_\perp}{J_\parallel}\right)_{cr}=\sqrt{2}\left[\sin^2(\pi n)-\cos^2(\pi n)\right].
\end{equation}
A special case occurs at the filling $n_0= \frac{1}{4}$ (and $\frac{3}{4}$), where an infinitesimal pump field amplitude is enough in order to generate a finite cavity field (see Fig.~\ref{fig:alpha_n_A} lower panel). At a filling $n<n_0$ ($n>1-n_0$) the occupation of the cavity field rises slowly above the pump strength $A_{cr}$, whereas for fillings $1/2>n>n_0$ a jump of the occupation of the cavity field to a finite value occurs at the critical pump strength $A_{cr}$ (see Fig.~\ref{fig:alpha_n_A} lower panel). 
At the same critical pump strength a finite chiral current of the atoms arises. 

\begin{figure}[hbtp]
\centering
\includegraphics[width=.5\textwidth]{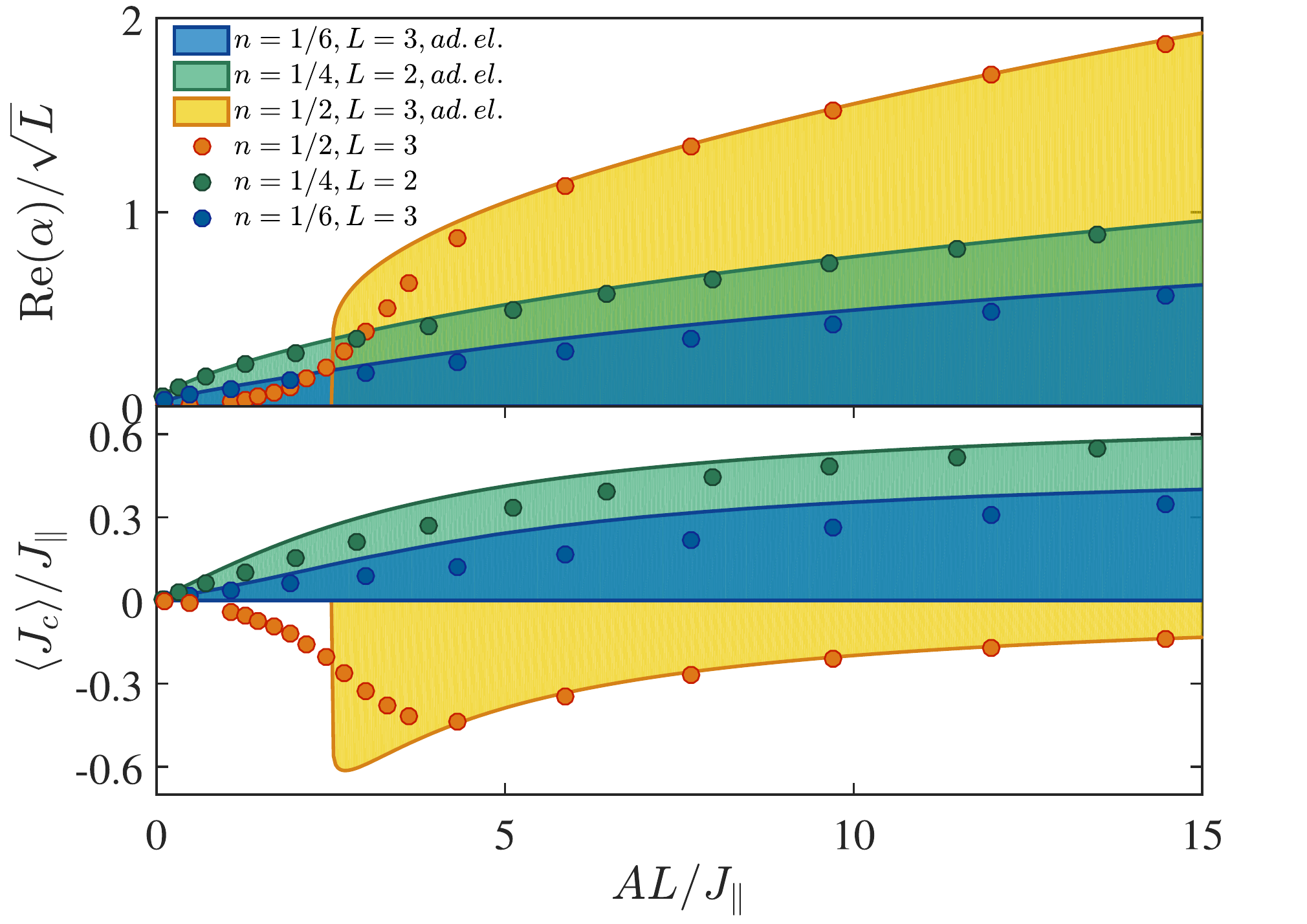}
\caption{Real part of the rescaled cavity field $\langle a\rangle$ (upper panel) and the expectation value of the chiral current $\langle J_c\rangle$ (lower panel) for different fillings $n$. The results from the adiabatic elimination (lines) of the cavity field for finite size $L$ are compared to the results from the exact diagonalization method (symbols). A finite current arises at finite cavity field amplitudes. The loss rate is $\hbar\kappa/J_\|=0.05$ and the cavity detuning is $\hbar \delta_{cp}/J_\| = 1.0$. The cutoff of the boson Fock space for the exact diagonalization is chosen to be $25$. }
\label{fig:alpha_Jc_A}
\end{figure}

The chiral current shows a more complex behavior with pump strength compared to the amplitude of the cavity field (Fig.~\ref{fig:alpha_Jc_A} lower panel). At half filling the chiral current jumps to a finite maximal value and then decreases in magnitude with increasing pump strength, whereas at low filling $n=1/4$ and $n=1/6$, it increases with the pump strength. 

In Fig.~\ref{fig:alpha_Jc_A} the results of the adiabatic elimination of the cavity field are compared to the data obtained by exact diagonalization of the master equation (Eq. \ref{eq:Lindblad}). The adiabatic elimination results are evaluated for the same system lengths reachable with the exact diagonalization method. The main behavior of the cavity field and the chiral current agrees between the results obtained by the two different approaches. In particular, the results for $n=1/6,1/4$ both show a slow rise of the cavity field amplitude and also of the chiral current. In contrast to that, for $n=1/2$ , the sudden jump of the occupation of the cavity field at $A_{cr}$ to a finite value is smoothed in the numerical results for the small systems. This might originate from the influence of fluctuations. Whereas the exact diagonalization suffers from finite system fluctuations and thus overestimates fluctuations, in the adiabatic elimination fluctuations are only partly taken into account as it neglects all quantum correlations between cavity photons and atoms. 

\begin{figure}[hbtp]
\centering
\includegraphics[width=0.5\textwidth]{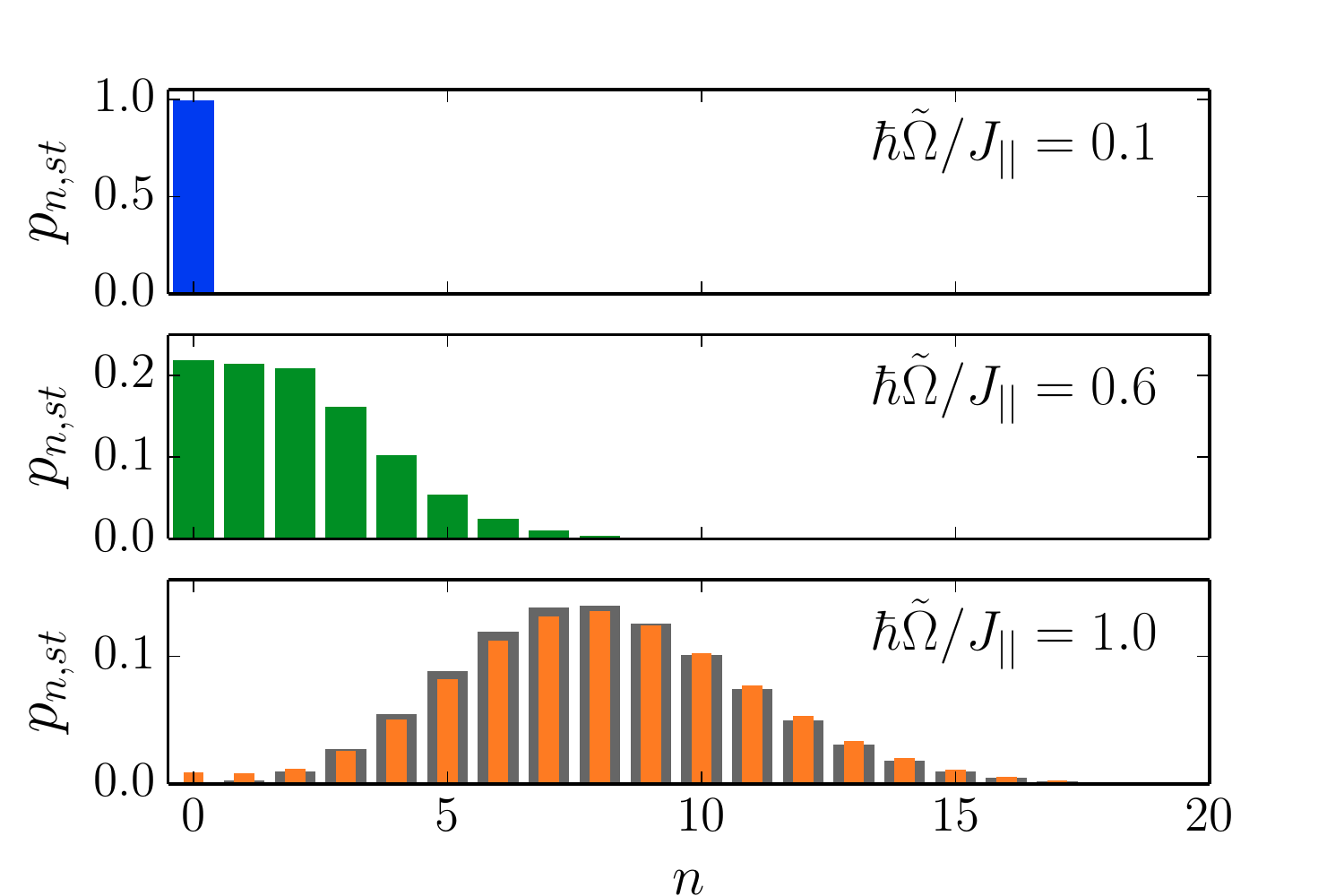}
\caption{Distribution of the cavity mode occupation in steady state $p_{n,st}$ for different pump strengths $\tilde\Omega$ calculated by the exact diagonalization method for $\hbar \kappa /J_\|=0.25$, $\hbar \delta_{cp}/J_\|=1.0$. $L=3$ and $n=0.5$. The bottom panel includes a Poissonian fit (grey) to the discrete distribution function.}
\label{fig:photon_distribution}
\end{figure}

The nature of the cavity field in the steady state can be best identified with the help of the photon number distribution (Fig. \ref{fig:photon_distribution}). With increasing pump strength the mean value of the cavity field distribution increases towards larger values. Additionally, a broadening of the distribution with increasing pump strength is found. Due to the finite and relatively small number of photons, the form of the distribution cannot be identified uniquely. It still complies with both a Gaussian and Poissonian distribution. In particular, for the shown distribution at $\hbar \tilde{\Omega}=J_\parallel$, the description by a Poissonian distribution has been tested to be within a confidence interval larger than 99\%.

\begin{figure}[hbtp]
\centering
\includegraphics[width=0.5\textwidth]{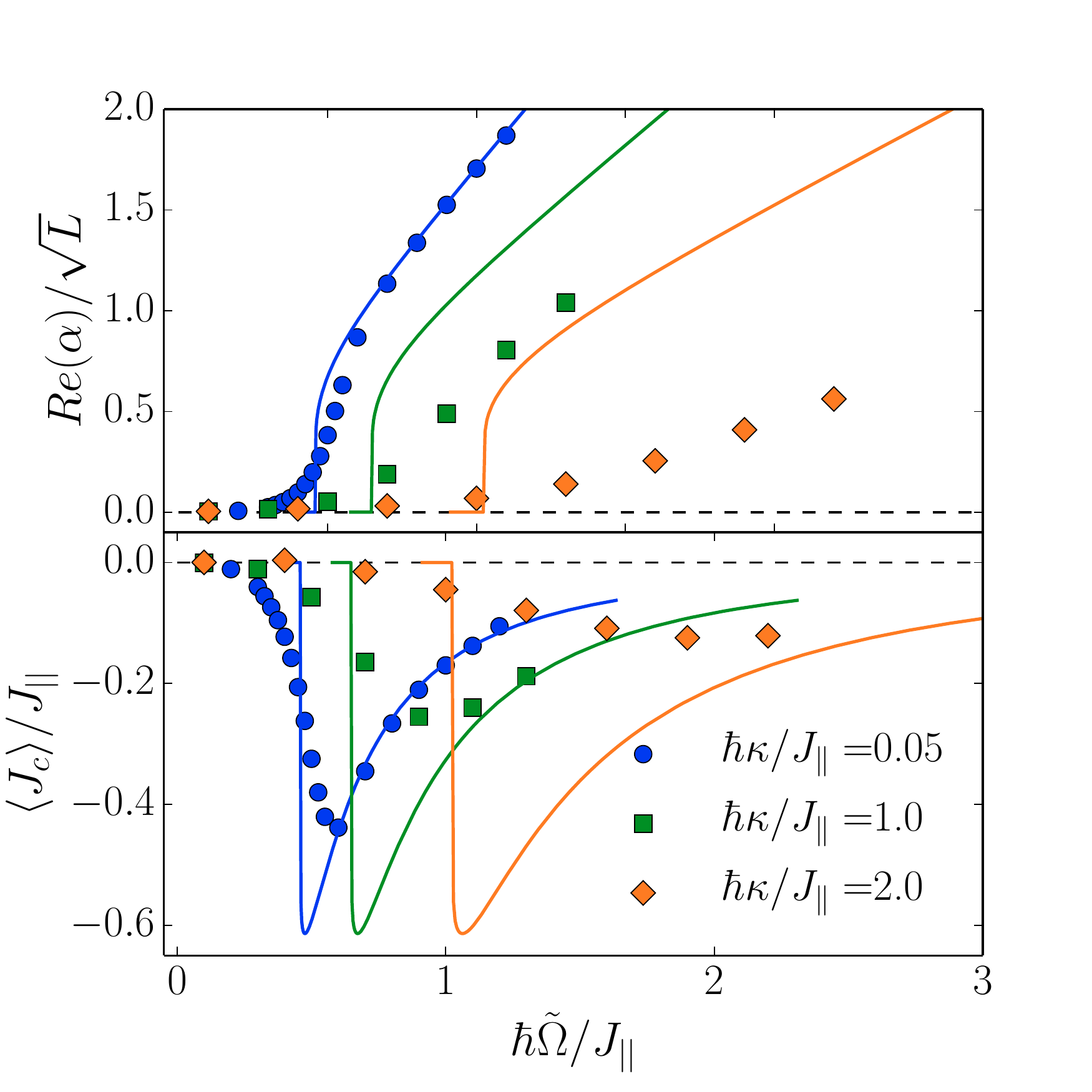}
\caption{The rescaled real part of the cavity field expectation value $Re (\alpha)$ (upper panel) and the chiral current $J_c$ (lower panel) versus the pump strength $\hbar\tilde \Omega/J_\|$ for different strengths of the photon loss rate $\hbar\kappa/J_\|$. Lines display the adiabatic elimination results  and symbols are the exact diagonalization results, both for a system at half filling with $L=3$ rungs and $N=3$ atoms. The data is obtained for $\hbar\delta_{cp}/J_\|=1.0$ and for the exact diagonalization the cutoff of the photon number is $20$. }
\label{fig:kappa}
\end{figure}

\begin{figure}[hbtp]
\centering
\includegraphics[width=0.5\textwidth]{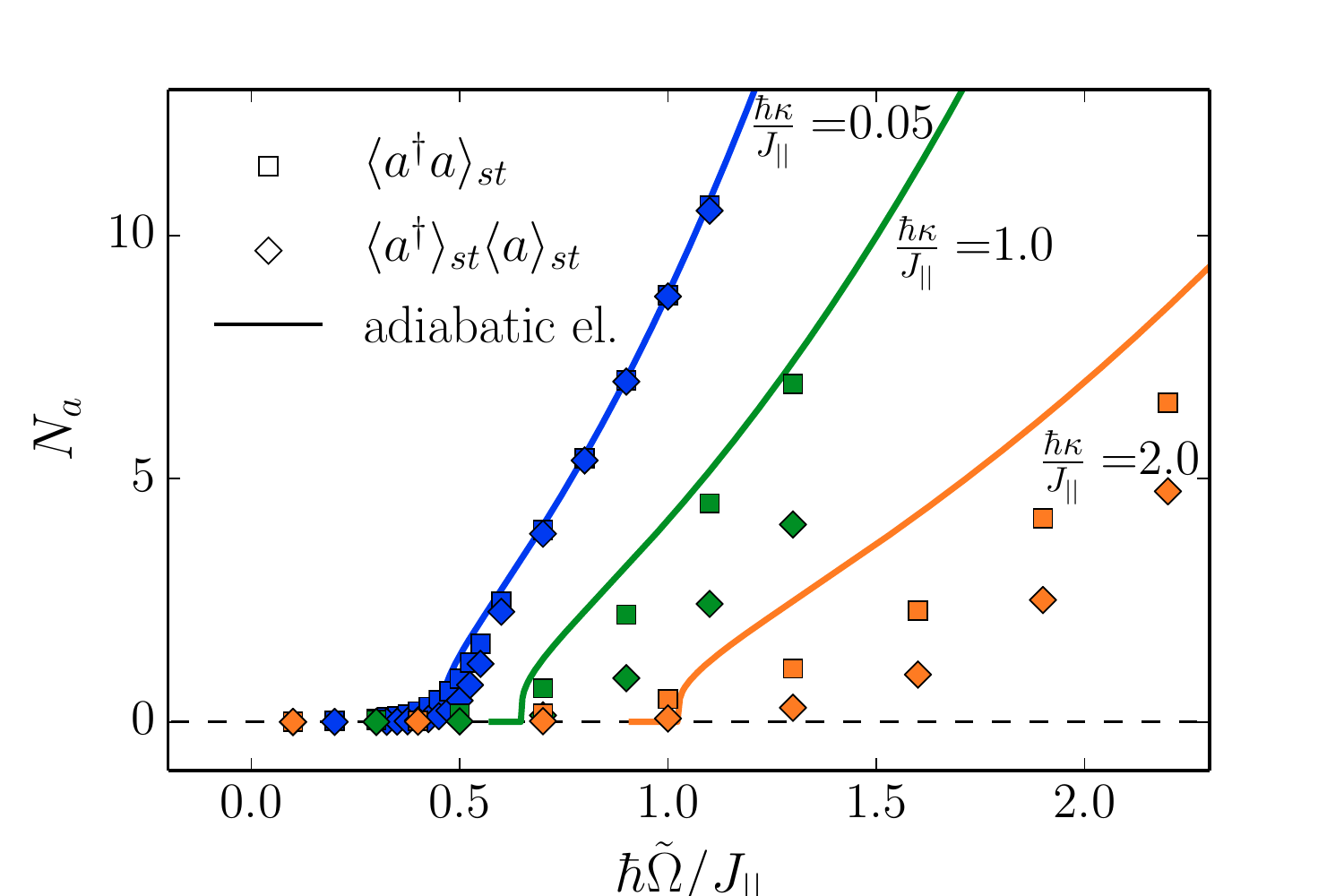}
\caption{Dependence of the cavity occupation $N_a$ on the pump strength measured directly $\langle a^\dagger a \rangle$ (squares) and using decoupled mean-fields, i.e.~$\langle a^\dagger\rangle \langle a\rangle=\vert \alpha\vert^2$ (diamonds) obtained by exact diagonalization (symbols) in comparison to finite-size adiabatic elimination results (lines) for different photon loss rates of $\hbar\kappa/J_\|=0.05$ (blue), $1.0$ (green) and $2.0$ (orange). Here we used $\hbar \delta_{cp}/J_\|=1.0$, $N=3$, and $L=3$.}
\label{fig:meanfield}
\end{figure}

In the upper panel of Fig.~\ref{fig:kappa}, we show the cavity field amplitude versus pump strength for different values of the photon loss rate $\kappa$ at half filling $n=0.5$. The effect of the dissipation introduces a smearing of the jump visible in the adiabatic elimination results. For a low value of $\kappa$ the numerically calculated cavity amplitude increases quickly at the position of the jump, whereas for a larger value of $\kappa$ only a slow increase is observed. Similarly, the onset of the chiral current in the lower panel of Fig.~\ref{fig:kappa} which is relatively steep for small values of $\kappa$ becomes much broader and its maximal magnitude drops drastically.  Part of the discrepancy between the exact diagonalization results and the adiabatic elimination at larger $\kappa$ stem from neglecting fluctuations. In Fig.~\ref{fig:meanfield} we show the comparison of the photon occupation number $N_a$ and its  mean-field decoupling into $\aver{a}\aver{a^\dagger}$ calculated with the exact diagonalization. Additionally, the results from adiabatic elimination are shown. As one can see the mean field decoupling deviates strongly from the exact results already for intermediate values of $\kappa$  due to the presence of fluctuations. This effect is expected to become smaller for larger system sizes.

\section{Dissipative dynamics after the quench}
\label{sec:dynamics}

\begin{figure}[hbtp]
\centering
\includegraphics[width=.5\textwidth]{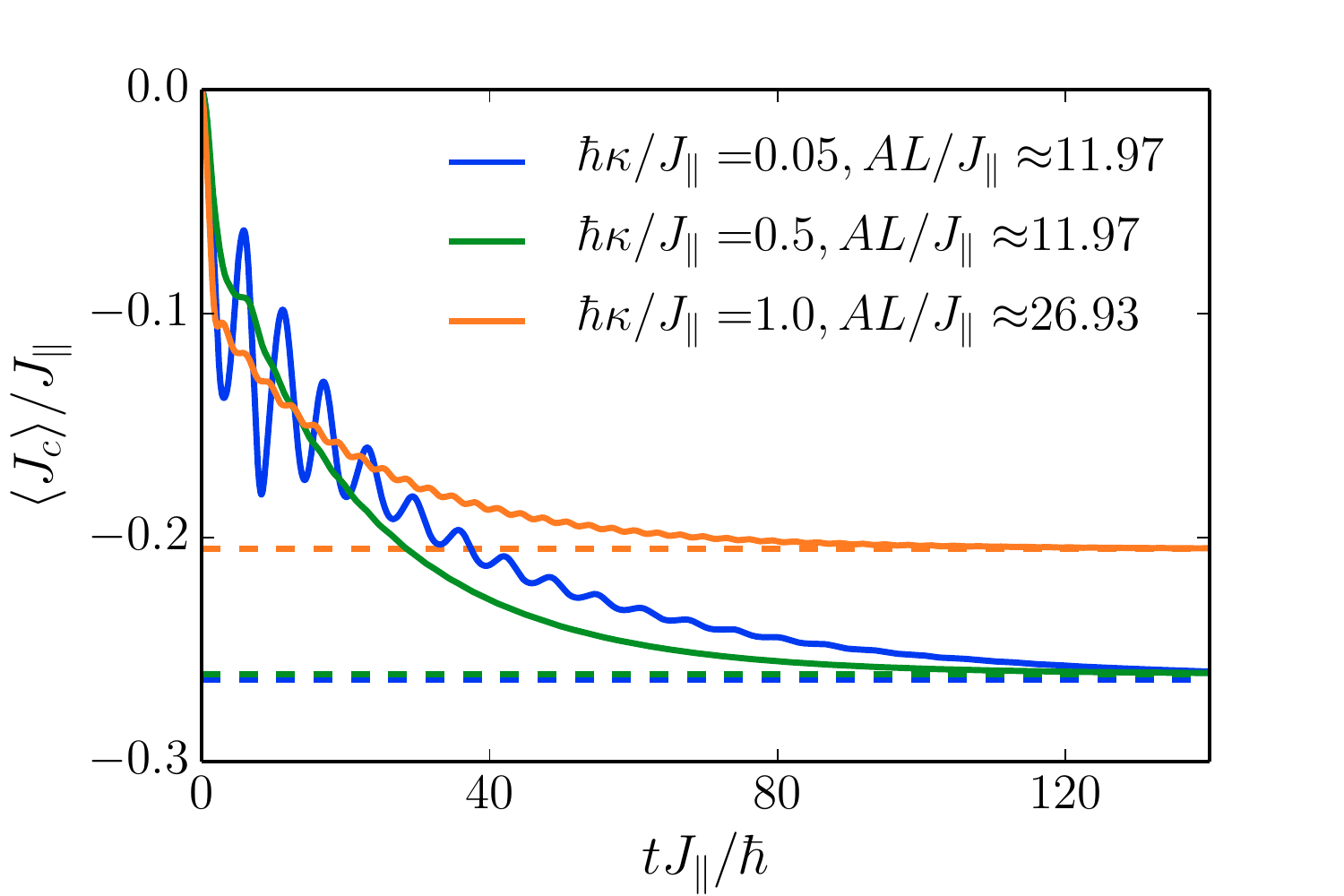}
\caption{Time evolution of the expectation value of the chiral current $\aver{J_c}$ for different parameter sets starting from the pure initial state $\ket{\psi_i}$. The corresponding steady states are indicated by dashed lines. The relaxation towards the steady states strongly depends on the atom losses $\kappa$. Here $L=3$, $N=3$ and $\hbar \delta_{cp}/J_\| = 1.0$. }
\label{fig:td_current}
\end{figure}

In this section we discuss the full time evolution of the system starting from a pure initial state. The fermionic part is positioned in one of the symmetry blocks of the fermionic many-body Hamiltonian as described in appendix \ref{app:symmetry_transformation} and the cavity mode is initially empty. For the chosen symmetry sector the chiral current has the maximum value and matches with the adiabatic elimination results. At time $t=0$ the coupling to the cavity mode is switched on. We are particularly interested in how the state approaches the steady state. An examplary evolution of the chiral current is shown in Fig.~\ref{fig:td_current} for different parameter sets. Starting from the abscence of a chiral current in all cases a fast decrease of the chiral current and a damped oscillating behavior is observed. At longer times the steady state values are approached (indicated by dashed lines). The value of the chiral current for different parameter sets but the same scaled pump strength $AL/J_\parallel$ are equal within the adiabatic elimination.  However, in the exact diagonalization slight deviations occur. The time-scale of the approach of the steady state strongly depends on the value of the dissipative coupling $\kappa$ as will be analyzed in more detail below. 

\begin{figure}[hbtp]
\centering
\includegraphics[width=0.5\textwidth]{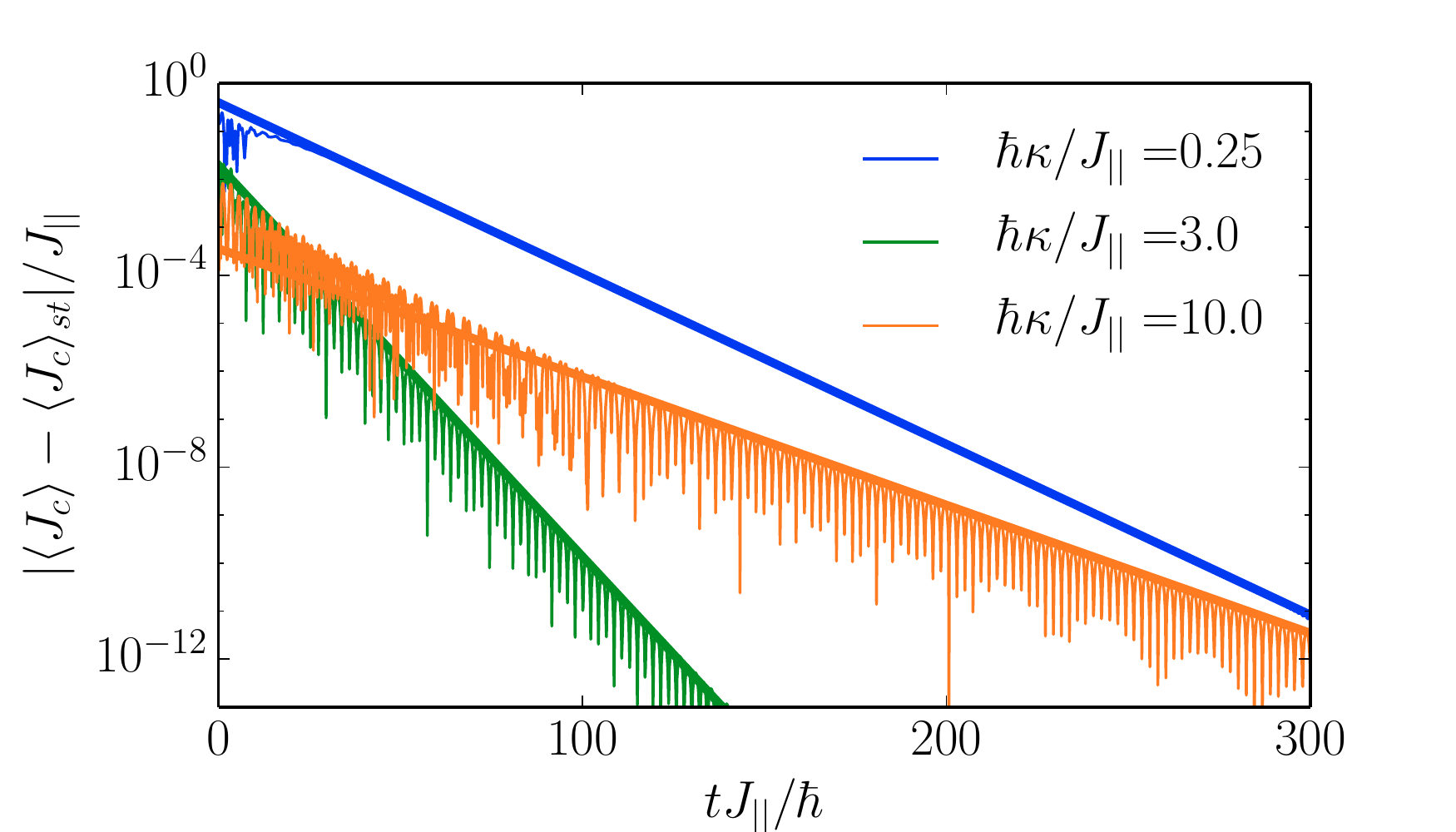}
\caption{Approach of the steady state expectation value of the chiral current in time for different values of $\hbar\kappa /J_\|$ with $\hbar\tilde\Omega/J_\| = 0.8$, $\hbar \delta_{cp}/J_\|=1.0$, $L=3$ and $N=3$. Straight lines are exponential fits. The fitted slopes in the logarithmic plot provide the negative inverse time-scale $-1/\tau$ of the relaxation process.}
\label{fig:exponentialapproach}
\end{figure}

\begin{figure}[hbtp]
\centering
\includegraphics[width=0.5\textwidth]{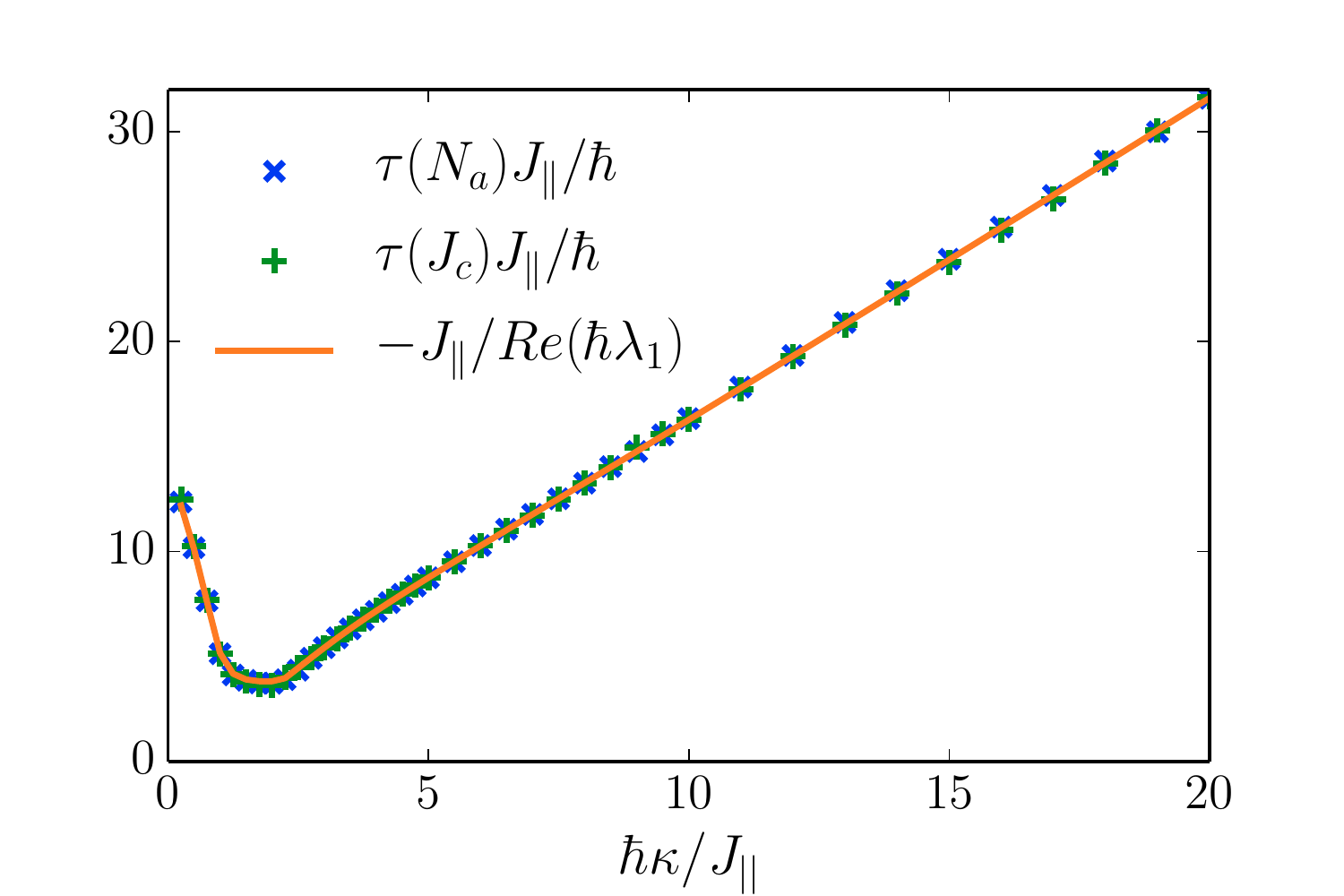}
\caption{Dependence of the time scales for the relaxation towards the steady state on $\hbar \kappa/J_\|$ extracted by exponential fits to the time evolution (see Fig.~\ref{fig:exponentialapproach}). The time-scales of the cavity occupation $\tau(N_a)$ (blue) and the chiral current $\tau(J_c)$ (green) coincide. Errors from the fit are smaller than the symbols. 
Both time scales are in agreement with the negative inverse of the lowest non-vanishing absolute real part of the eigenspectrum $\lambda_1$ shown by the solid line, which is associated with the slowest dynamics. We chose $\hbar \tilde \Omega/J_\|=0.8$, $\hbar\delta_{cp}/J_\|=1.0$, $L=3$ and $N=3$.}
\label{fig:timescales}
\end{figure}

In order to get more insights in the approach of the steady state value we exhibit in Fig.~\ref{fig:exponentialapproach} the time evolution of the difference between the chiral current and its steady state value. A clear exponential decay of this difference is found at late times. On top of this exponential decay oscillations with decreasing amplitude are observed. Already from this plot it is clear that the time-scale of the exponential decay first decreases with increasing $\kappa$ and then increases again. 

\begin{figure}[hbtp]
\centering
\includegraphics[width=.5\textwidth]{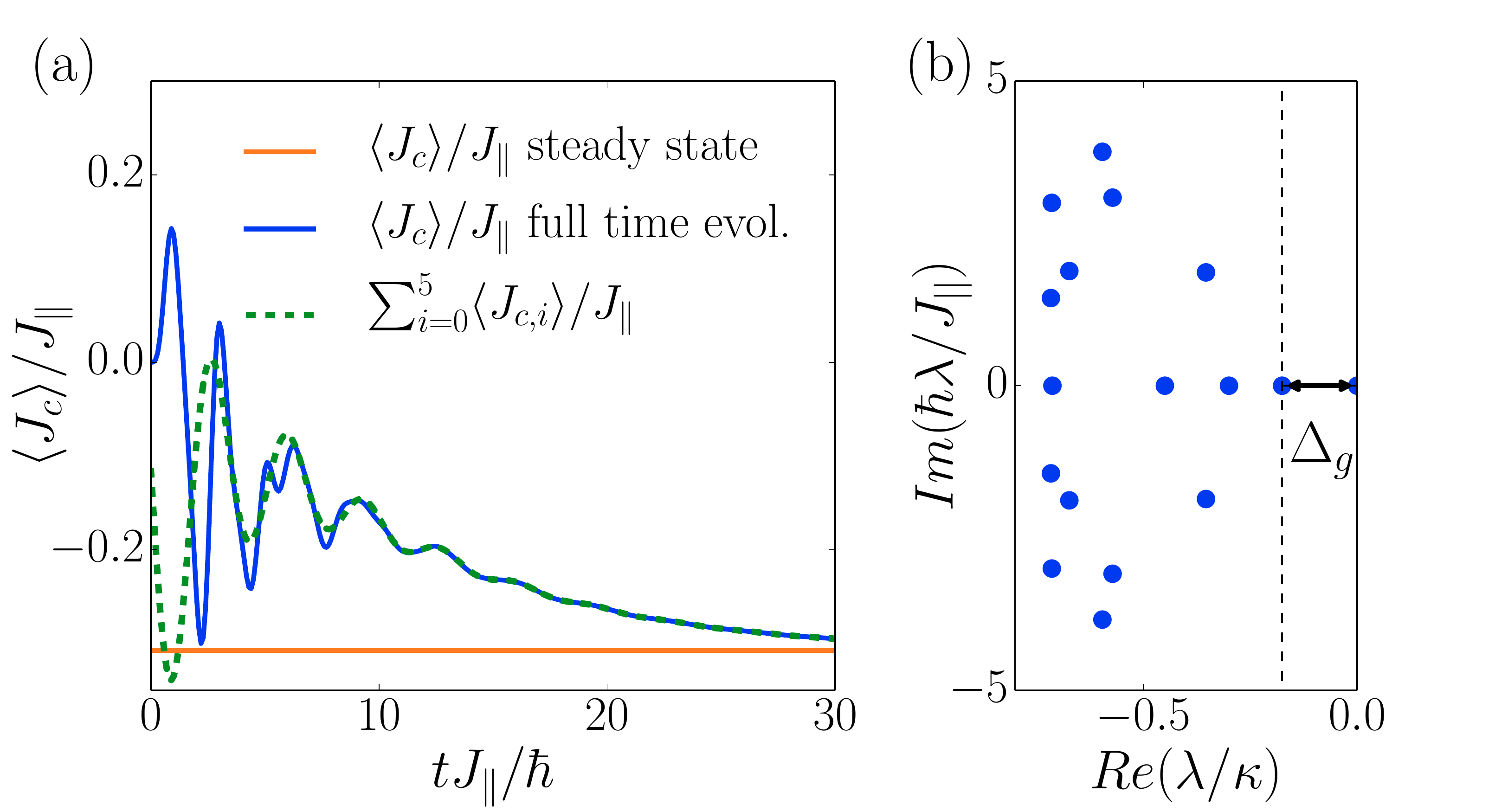}
\caption{(a) Time evolution of the expectation value of the chiral current $\aver{J_c}$. The dashed line represents the cumulated current of the eigenstates corresponding to the six eigenvalues with the lowest absolute real part, including the steady state. (b) Eigenspectrum of the Lindblad matrix in the complex plane showing the eigenvalues with the lowest absolute real parts. Only the eigenvalues in the relevant symmetry block for the chosen initial state are considered. The parameters are chosen as $\hbar \kappa/J_\| = 5.0$, $\hbar \delta_{cp}/J_\|=1.0$, $L=3$ and $N=3$, and $\hbar \tilde \Omega/J_\|=0.8$. The spectral gap is labeled as $\Delta_g$.}
\label{fig:jcdomination16loweststates}
\end{figure}

Extracting the time-scale of the steady state approach for the chiral current and  the cavity field occupation, one finds that both the fermionic time scale and the cavity time scale lie on top of each other (Fig.~\ref{fig:timescales}).
In addition to the time-scale extracted from the time-dependent simulations, we show the time-scale corresponding to the lowest non-vanishing magnitude of the real part of the eigenvalues of the relevant symmetry block for the chosen initial state in Fig.~\ref{fig:timescales}. A very good agreement between the different time-scales is found. This is due to the fact that the decay of the remaining eigenvalues contribution is fast compared to the long time-scale set by the spectral gap $\Delta_g$ which is defined by the lowest non-vanishing absolute real part of the eigenvalues in the spectrum (Fig.~\ref{fig:jcdomination16loweststates}~(b)). In some cases this lowest absolute real part is shared among several eigenvalues. To confirm this finding we show in Fig.~\ref{fig:jcdomination16loweststates} (a) the full time evolution and compare it to the evolution taking only the six eigenvalues with the lowest absolute real part into account. The fast initial dynamics is very distinct, however, already after a short time of the order of $\hbar/J_\parallel$ the evolutions agree well.

\begin{figure}[hbtp]
\centering
\includegraphics[width=0.5\textwidth]{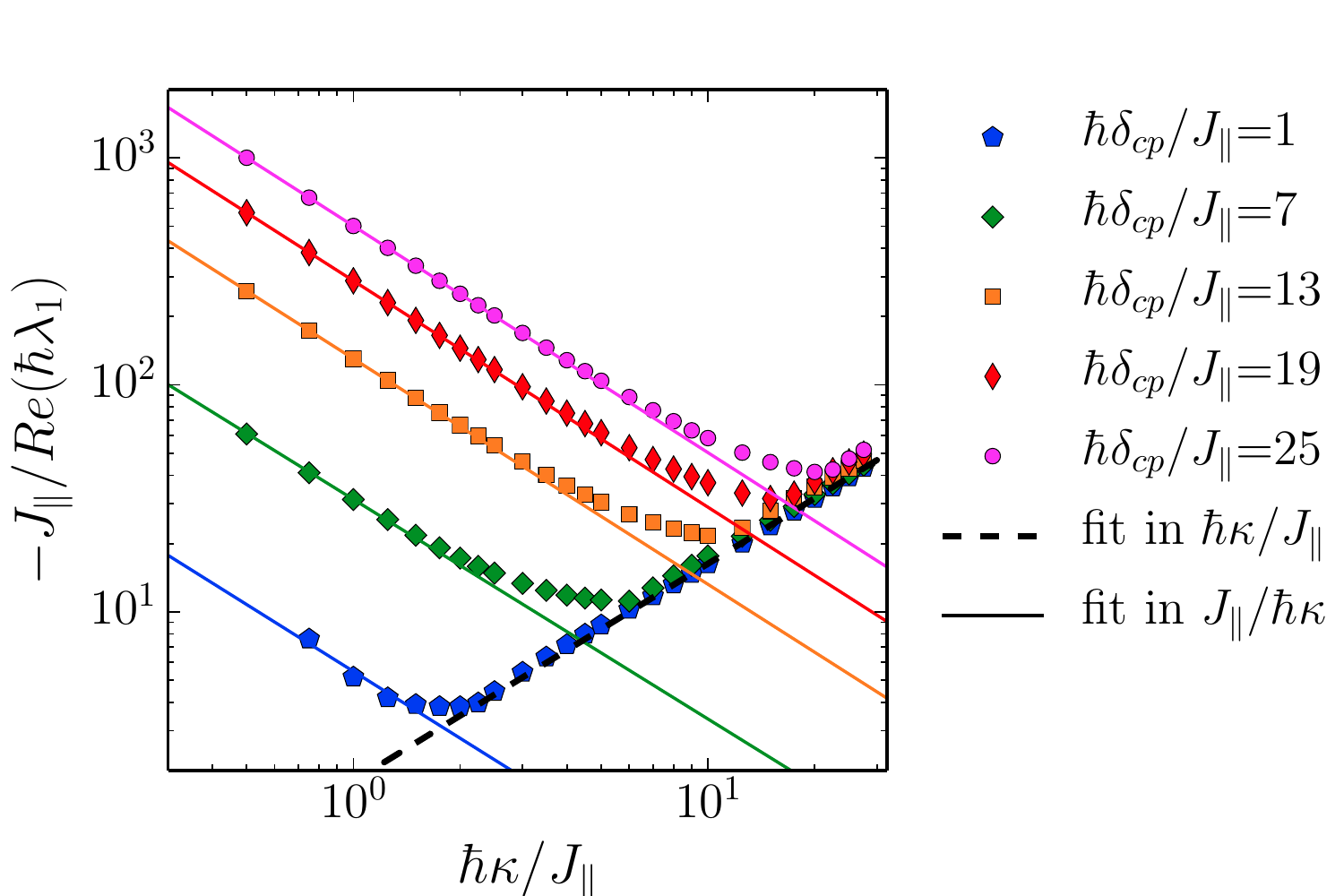}
\caption{Negative inverse of the lowest non-vanishing absolute real part of the spectrum of the Linblad operator for different rotating cavity frequencies $\delta_{cp}$ versus the cavity losses $\hbar\kappa/J_\|$. Solid lines are fits proportional to $J_\|/\hbar\kappa$ while the dashed black line corresponds to a fit proportional to $\hbar\kappa/J_\|$. Here $L=3$, $N=3$ and $\hbar \tilde \Omega /J_\| =0.8$.}
\label{fig:ts_diffdelta}
\end{figure}

As seen in Fig.~\ref{fig:timescales} the time-scale of the decay towards the steady state depends on the dissipation $\kappa$. At low values of $\kappa$ the time scale becomes shorter with increasing values of $\kappa$. This is as naively expected. For larger loss rate the photons, the steady state value is reached in a shorter time. More precisely, we find that the time-scale decreases as $1/\kappa$ (see Fig. \ref{fig:ts_diffdelta}). However, the decay time reaches a minimum and rises afterwards. This indicates the counterintuitive behavior, known as the Zeno effect, that for stronger dissipative coupling the system evolves slower towards the steady state. The corresponding time-scale is proportional to $\kappa$. A simple picture for this is that the continuous measurement process effectively introduced by the dissipative process freezes the system to its state. 

The limits of the time-scale for large and small values of $\kappa$ are already present for a single fermions on an isolated rung coupled to the cavity field. In this case the Hamiltonian reduces to $H_{\text rung}=\hbar\delta_{cp}a^\dagger a -\hbar\tilde{\Omega}(a+a^\dagger)(c_0^\dagger c_1 + c_1^\dagger c_0)$. The fermionic part of the Hamiltonian ($H_{{\text rung},f}=c_0^\dagger c_1 + c_1^\dagger c_0$) can be diagonalized with the two eigenvalues $\pm 1$. Thus, the Lindblad matrix $M_\mathcal{L}$ is block diagonal with four blocks that can be solved exactly for low cutoffs of the bosonic Fock space. The spectral gap considering only a photon occupation  of 0 or 1 and $\delta_{cp}=0$ corresponds to the eigenvalue $\lambda_1=\frac{1}{2}\left(-\kappa+\sqrt{\kappa^2-16\tilde{\Omega}^2}\right)$ which is a complex value in general. The time scale $\tau=-1/Re(\lambda_1)$ behaves as $1/\kappa$ and $\kappa$ for small and large dissipation constants $\kappa$, respectively. For non-zero $\delta_{cp}$ the eigenvalue has a more complex form but the behavior in both limits is the same.
 We show the limiting cases for different values of the parameter $ \delta_{cp}$ in Fig.~\ref{fig:ts_diffdelta}. The time scale for low values of $\kappa$ becomes longer with increasing values of $\delta_{cp}$. At larger values of $\kappa$ the curves approach each other and the proportionality to $\kappa$ is clearly seen. 

\section{Discussion}
\label{sec:discussion}
In this work we investigated the dissipative dynamics of fermionic atoms placed in an optical cavity. The fermionic atoms were confined by external optical lattices to ladder structures. The tunneling on the rungs is induced by a cavity assisted Raman process which allows a position dependent phase transfer. By the feedback between the cavity mode and the atoms, the cavity field amplitude becomes finite and gives rise to an artificial magnetic field felt by the atoms. This in turn induces a chiral current for the fermionic atoms \cite{KollathBrennecke2016,SheikhanKollath2016}. In this work we detailed the properties of the steady state and the dynamics after a coupling quench towards this steady state using an exact diagonalization of the master equation. The numerical results for the chiral current and occupation of the photons at the steady state are compared with the analytical results from the adiabatic elimination of the cavity field. They agree well for low cavity loss. We find that for stronger coupling strength the transition to the chiral state with the pump strength is broadened considerably. This is related to the quantum fluctuations in the system. 

The time evolution after the quench towards the steady states is characterized by a dissipative attractor dynamics. Starting with an initial state with an empty cavity, the chiral state is reached exponentially fast with a time-scale depending strongly on the dissipative coupling $\kappa$. Whereas at low values of $\kappa$ the dynamics becomes faster as $1/\kappa$ with increasing values of $\kappa$ as intuitively expected, above a certain value the contrary occurs and the dynamics slows down proportional to $\kappa$. This is related to the Zeno effect and we analyzed in detail the dependence of the time-scale.

The presented setup is an example for the dynamic transition from a state without chiral current to a state with chiral current. Such a transition has been investigated recently a lot in the context of topologically non-trivial states, however, mainly in isolated systems \cite{RahmaniChamon2010,Perfetto2013,HalaszHamma2013,TsomokosFazio2009,FosterYuzbashyan2013,WangXianlong2015,DongPu2014b,FosterYuzbashyan2014,PatelDutta2013,WangKehrein2016,DongPu2015,DAlessioRigol2015,WangKehrein2015,CaioBhaseen2015}. In contrast to the isolated systems, in this work the presence of the cavity photon losses leads to the exponentially fast stabilization of the chiral state by the attractive nature of the dynamics. The discussed setup is therefore an important example how to engineer chiral states in a fast and robust way using dissipative processes very much in the sense of open system control \cite{MuellerZoller2012}.

\begin{acknowledgments}
We would like to thank F. Brennecke for useful suggestions and careful reading of the manuscript. We acknowledge fruitful discussions with J.-S.~Bernier, F.~Piazza, H.~Ritsch, and W.~Zwerger and support from the DFG (among others FOR1807) and the ERC (Grant Number 648166, Phon(t)on). 
\end{acknowledgments}

\appendix
\section{Symmetry transformation of the fermionic sector}
\label{app:symmetry_transformation}
Here we present the symmetry transformation to block diagonalize the fermionic part of the single particle Hamiltonian. 
This reduces the numerical complexity of the problem significantly and combined with many body symmetries gives direct access to the steady states which are distributed over the resulting symmetry blocks. The $S_1$ symmetry transformation explained in the text (see Eq. \ref{eq:symm}) is given by
\begin{align}
S_1 =& \sum_{j =0}^{L-1} e^{-i\frac{L-1}{2}\varphi} c_{1,L-1-j}^\dagger c_{0,j} + \sum_{j =0}^{L-1} e^{+i\frac{L-1}{2}\varphi} c_{0,L-1-j}^\dagger c_{1,j}.
\end{align}
The single particle basis for the two-leg ladder with $L$ rungs is given by $\{\vert n\rangle = \vert m\cdot L + j\rangle = c_{m,j}^ \dagger\vert 0\rangle\}$, where $m\in\{0,1\}$ and $j\in\{0,1,\ldots,L-1\}$ represent the chain and rung index, respectively. 

We introduce the operators 
\begin{align}
\gamma_{\tilde n}^\dagger=\frac{1}{\sqrt{2}}\times\begin{cases}
-\text{e}^{+i\frac{L-1}{2}\varphi} c_{0,\tilde n}^\dagger +  c_{1,L-\tilde n -1}^\dagger &, \tilde n \in  \{0,1,\ldots, L-1\}\\
+\text{e}^{+i\frac{L-1}{2}\varphi} c_{0,\tilde n - L}^\dagger +  c_{1, 2L-\tilde n -1 }^\dagger &, \tilde n \in  \{L,\ldots, 2L-1\} .
 \end{cases}
\end{align}
A new single particle eigenbasis is spanned by 
 $\vert \tilde n \rangle =\gamma_{\tilde n}^\dagger \vert 0 \rangle $. Here the vectors $\ket{ \tilde n }$ with $ \tilde n\in \{0,1,\ldots, L-1\} $ lie in the symmetry block which corresponds to the Hilbert space spanned by the eigenvectors of $S_1$ corresponding to the eigenvalue $-1$.
Analogously the vectors $\ket{ \tilde n }$ with $ \tilde n\in \{L,\ldots, 2L-1\} $ lie in the symmetry block which corresponds to the Hilbert space spanned by the eigenvectors of $S_1$ corresponding to the eigenvalue $1$.

In the basis of $\ket{ \tilde n }$ the fermionic parts of the single particle Hamiltonian are block diagonal. More particularly, we find
\begin{align}
 H_\|/J_\| =& 
 - \sum_{\tilde n =0}^{L-2}  \left( \gamma_{\tilde n}^\dagger\gamma_{\tilde n +1} + \gamma_{\tilde n +1 }^\dagger\gamma_{\tilde n }\right)\notag\\
&- \sum_{\tilde n =L}^{2L-2} \left( \gamma_{\tilde n}^\dagger\gamma_{\tilde n +1} + \gamma_{\tilde n +1 }^\dagger\gamma_{\tilde n }\right),\\
 H_\perp/J_\perp =& 
 + \sum_{\tilde n=0}^{L-1} \text{e}^{-i \left(\frac{L-1}{2} -\tilde n\right)\varphi}  \gamma_{\tilde n}^\dagger \gamma_{L-\tilde n -1} \notag\\
&- \sum_{\tilde n=L}^{2L-1}\text{e}^{-i \left(\frac{3L-1}{2} -\tilde n \right)\varphi} \gamma_{\tilde n}^\dagger \gamma_{3L-\tilde n -1}
\end{align}
where for both $H_\|$ and $H_\perp$ the first and second sum represent the first and second symmetry block, respectively.

 The many body basis is $\vert \tilde n_1, \tilde n_2,\ldots,\tilde n_N \rangle = \gamma_{\tilde n_N}^\dagger \ldots \gamma_{\tilde n_2}^\dagger \gamma_{\tilde n_1}^\dagger\vert 0 \rangle$ with $\tilde n_1<\tilde n_2 < \ldots < \tilde n_N$ at filling $\frac{N}{2L}$. Writing the many body Hamiltonian in this basis yields a block-diagonal form. Further, we can apply additional Bogoliubov transformations to split up the matrix blocks in even smaller sectors, containing only one steady state each. 
As shown in Table (\ref{tab:numbersteadystates}) for $L=3$ at half filling ($N=L$) there are $6$ blocks. The initial state of fermions is chosen as a pure state \begin{equation}
\ket{\psi_{t=0}}_{\text{Fermion}} = \vert \tilde 0 \tilde 2 \tilde 4 \rangle,
\end{equation}
which is located in one of the symmetry sectors.
As mentioned before, the cavity mode is initially unoccupied such that the initial state is given by $\ket{\psi(t=0)}=\ket{\psi_{t=0}}_{\text{Fermion}} \otimes \ket{N_a(t=0) =0} $.

\end{document}